\documentclass{iau}
\usepackage{graphicx}

\title[Late B- and early A-type periodic variables] 
{On the new late B- and early A-type periodic variable stars}

\author[N. Mowlavi, S. Saesen, F. Barblan \& L. Eyer]   
{Nami Mowlavi,
 Sophie Saesen,
 Fabio Barblan
 \and Laurent Eyer}

\affiliation{Geneva Observatory, University of Geneva, \\
51 chemin des Maillettes, 1290 Versoix, Switzerland
\\email: {\tt Nami.Mowlavi@unige.ch}}

\pubyear{2013}
\volume{301}  
\pagerange{1--4}
\jname{Precision Asteroseismology}
\editors{J.A. Guzik, W.J. Chaplin, G. Handler \& A. Pigulski, eds.}

\begin{document}

\maketitle

\begin{abstract}
We summarize the properties of the new periodic, small amplitude, variable stars recently discovered in the open cluster NGC~3766.
They are located in the region of the Hertzsprung-Russell diagram between $\delta$~Sct and slowly pulsating B stars, a region where no sustained pulsation is predicted by standard models.
The origin of their periodic variability is currently unknown.
We also discuss how the Gaia mission, to be launched at the end of 2013, can contribute to our knowledge of those stars.
\end{abstract}


\section{Introduction}
\label{Sect:introduction}

Stars are known to pulsate if certain conditions are met in their interior.
A pulsation mechanism must be active which, for main-sequence stars, is the $\kappa$ mechanism acting on H and He ($\delta$~Sct stars), the $\kappa$ mechanism acting on the iron-group elements ($\beta$~Cep, slowly pulsating B -- or SPB -- stars, and most probably rapidly oscillating Ap stars), the `convective blocking' mechanism ($\gamma$~Dor stars), or turbulence in the outer convection zone (solar-type oscillations).
Stringent conditions must be met for those pulsation mechanisms to operate, which translate in the existence of `instability regions' in the Hertzsprung-Russell (HR) diagram, specific to each type of pulsating star (\cite[Pamyatnykh 1999]{Pamyatnykh99}, \cite[Christensen-Dalsgaard 2004]{Christensen-Dalsgaard04}).
On the main sequence, solar-like pulsators, $\gamma$~Dor stars, $\delta$~Sct stars, SPB stars and $\beta$~Cep stars form an almost continuous sequence with increasing luminosity, except at luminosities between $\delta$~Sct stars and SPB stars where no pulsator is expected to be found, neither from model predictions, nor from observations, at least until recently.
This `gap' is clearly seen in, for example, Fig.~3 of \cite[Pamyatnykh (1999)]{Pamyatnykh99} or in Fig.~22 of \cite[Christensen-Dalsgaard (2004)]{Christensen-Dalsgaard04}.

Recently, we have completed the analysis of photometric data gathered on NGC~3766 following a 7-year observation campaign, with the aim of reaching a census on the content of periodic variables in the $11.5' \times 11.5'$ field of view centered on the open cluster (\cite[Mowlavi et al.~2013]{Mowlavi13}, MBSE13 hereafter).
The obvious advantages of studying variable stars in a cluster are related to, among other properties, their being co-eval and at the same distance from the Sun.
As a result, the knowledge of their magnitudes enables a direct inference of their relative positions in the H-R diagram.
We found $\gamma$~Dor, $\delta$~Sct and SPB stars at the expected magnitude and frequency ranges.
But we also found, quite unexpectedly, a population of periodic variables in the magnitude range between those of $\delta$~Sct and SPB stars, at the milli-magnitude level of variability (see MBSE13 for details).

We summarize the properties of these new variables in Sect.~\ref{Sect:properties}, and briefly discuss the origin of their variability in Sect.~\ref{Sect:origin}.
The potential contribution of the Gaia mission to our understanding of those stars is given in Sect.~\ref{Sect:gaia}.
Conclusions are drawn in Sect.~\ref{Sect:conclusions}.

\section{Properties of the new class of late B- and early A-type variables}
\label{Sect:properties}

\begin{figure}[t]
\begin{center}
 \includegraphics[width=2.4in]{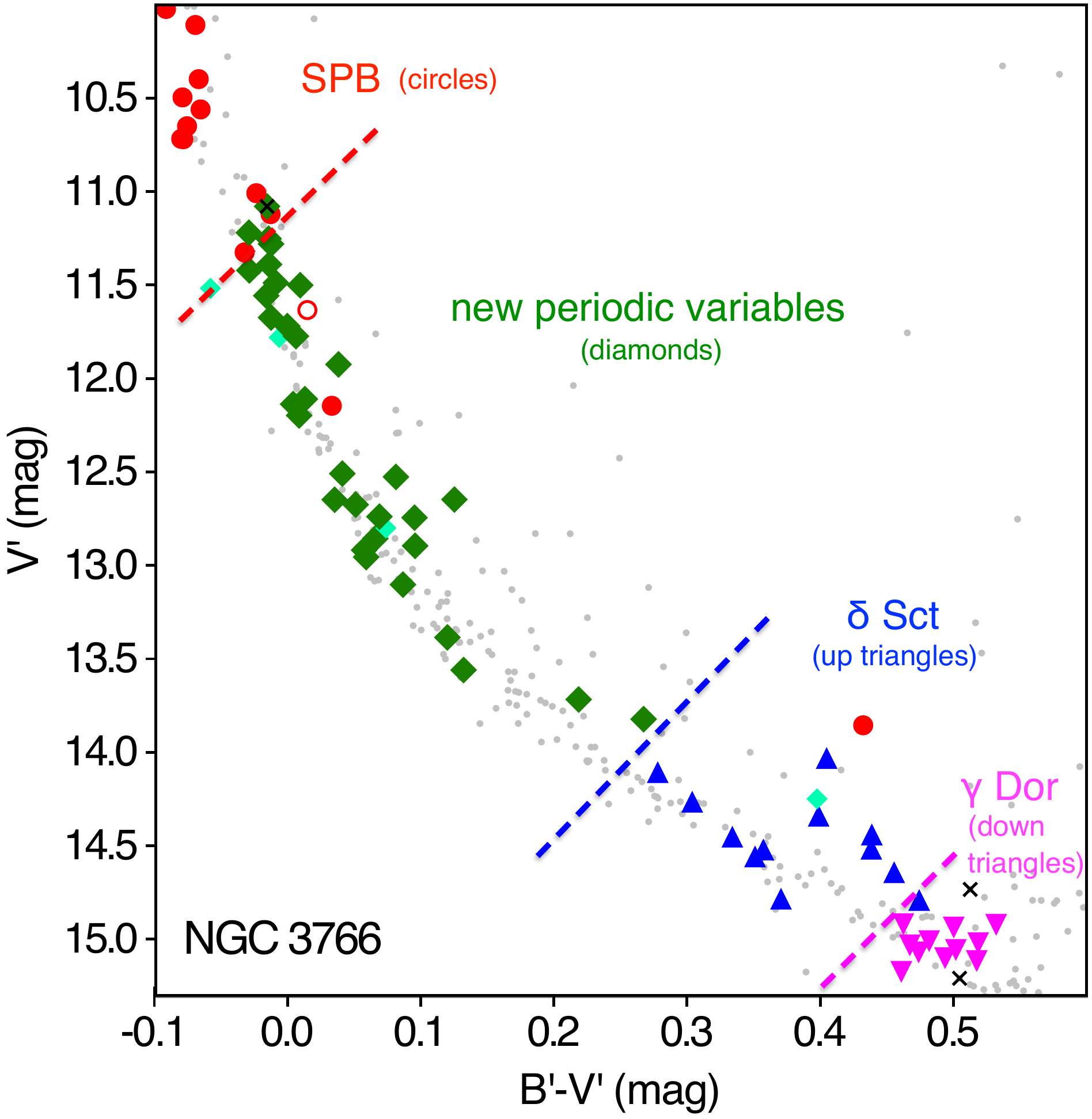}
 \includegraphics[width=2.4in]{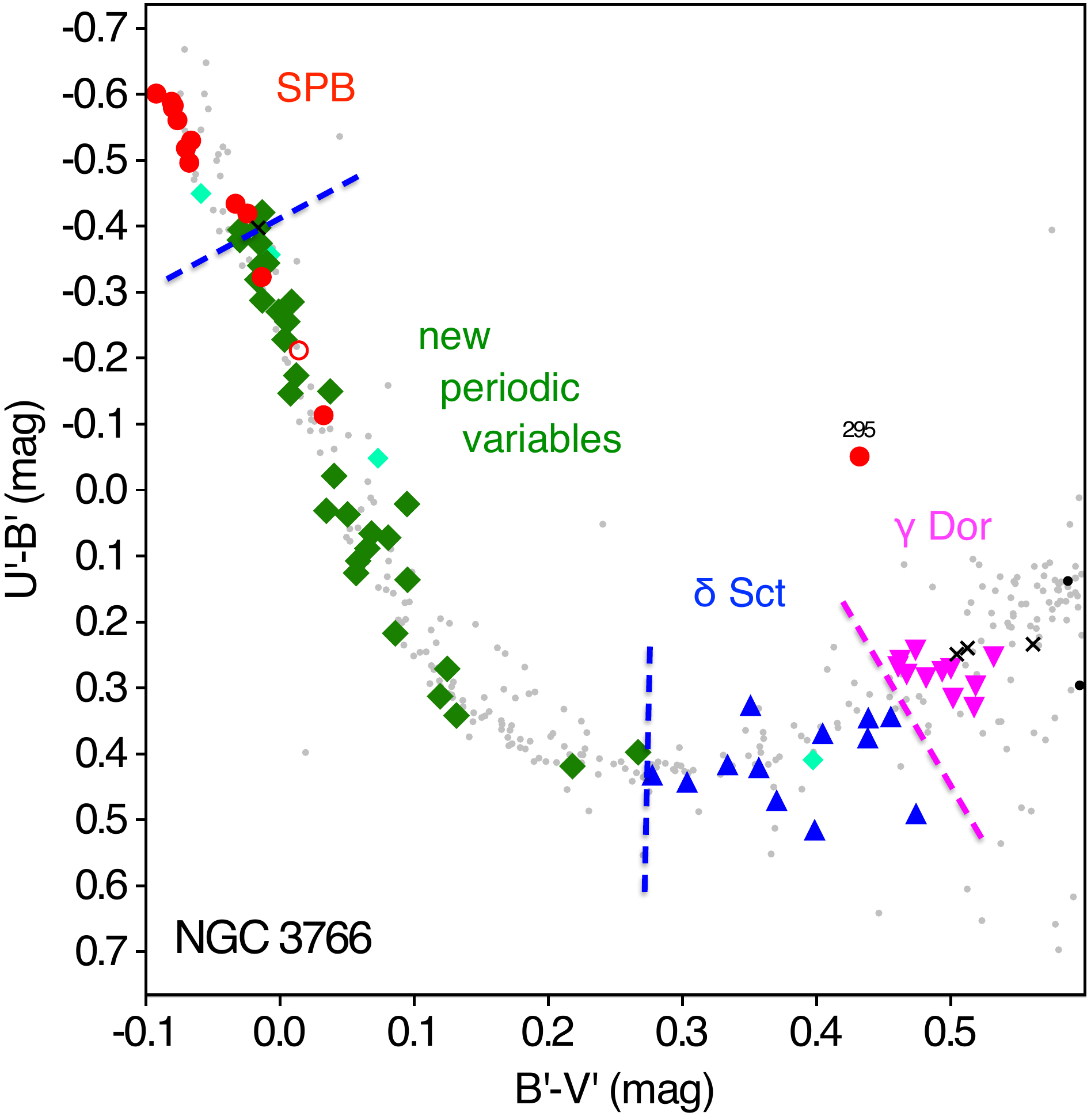}
 \caption{Distribution in the color-magnitude (left figure) and color-color (right figure) diagrams of main-sequence periodic variables detected in NGC~3766 in the magnitude range between those of $\gamma$~Dor and SPB stars.
Filled colored circles represent periodic variables, with the color and marker type identifying the different groups of variables as indicated on the figure.
Black crosses identify eclipsing binaries.
Grey dots represent constant or non-periodic variables.
The dashed lines separate the regions in the diagrams where the different groups of variables, identified from our data, are found.
}
\label{Fig:hr}
\end{center}
\end{figure}

The new periodic variables discovered in NGC~3766 have periods between 0.1 and 0.7~days and amplitudes of variability below few mmag (see Fig.~19 in MBSE13; four stars have periods up to 1.1~d, but they may be outliers).
The low amplitudes would explain why they are not easily detectable from ground.
We refer to Sect.~7.1 of MBSE13 for a summary of all the properties of those variables.

The striking characteristic of these variables is the fact that they lie outside the predicted instability strips in the H-R diagram of both SPB and $\delta$~Sct stars (see Fig.~\ref{Fig:hr}).
In addition, their periods differ from the typical periods expected for those two groups.
While SPB stars have periods between 0.5 and 5~d and $\delta$~Sct stars between 0.02 and 0.25~d, the majority of our new group of variables have periods in the range 0.25 -- 0.5~d.
Those two properties by themselves make those variables very peculiar, suggesting a new class of periodic variable stars.
The facts that their periods are stable over seven years (the duration of the observation campaign) and that up to one third of them are multiperiodic are further clues to understand the origin of their variability.

\begin{figure}[t]
\begin{center}
 \includegraphics[width=2.8in]{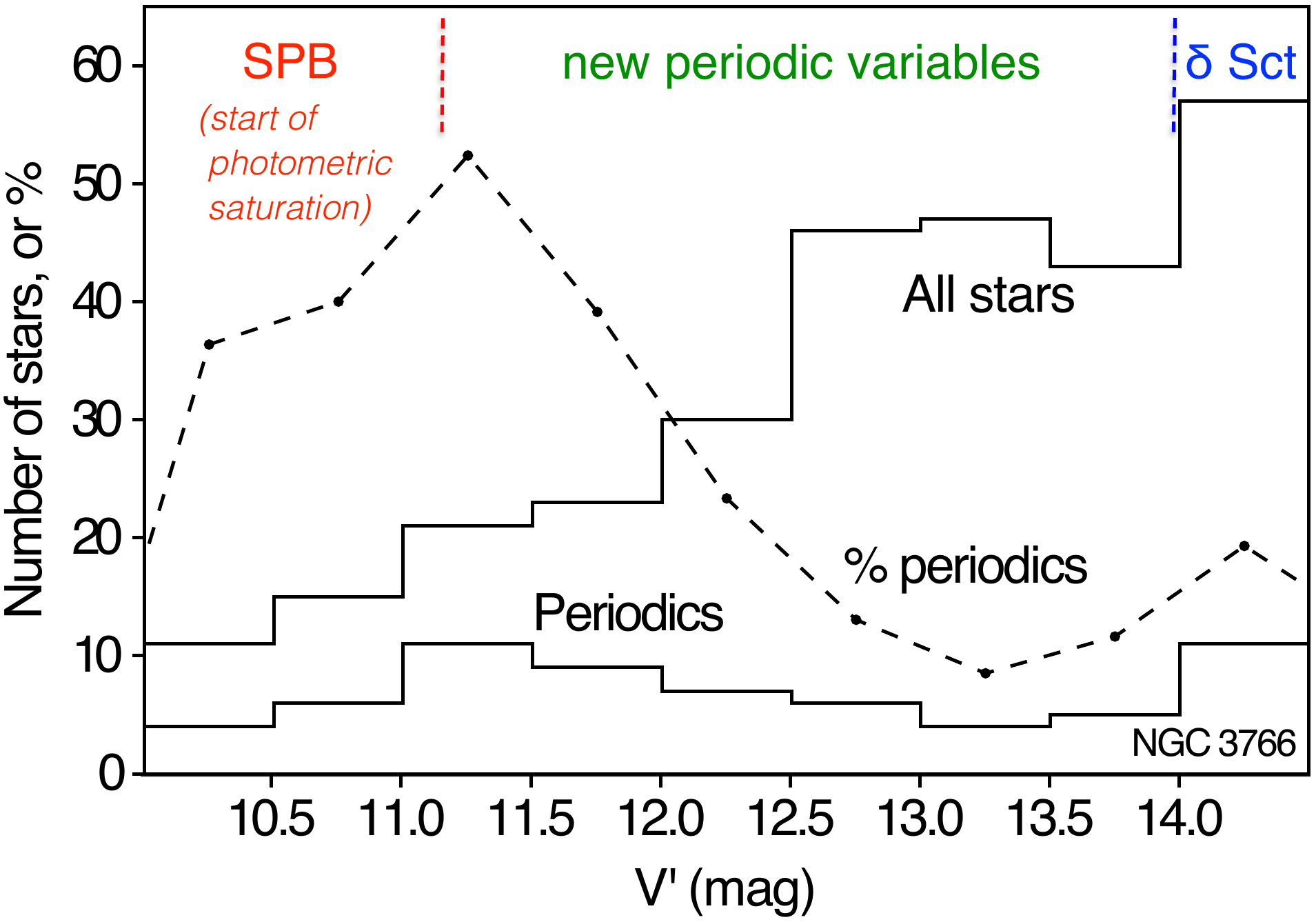}
 \caption{Total number of stars (upper histogram) and number of periodic stars (lower histogram) in NGC~3766 as a function of magnitude.
Only stars with $B'-V'<0.4$~mag are considered to exclude non-members of the cluster.
Markers connected with the dashed line give the percentage of periodic stars per magnitude bin.
The sharp decrease of the fraction of periodic variables at magnitudes below 11~mag is most probably due to photometric saturation in our data at those high luminosities.
The magnitude ranges where the different groups of variable stars are found in our data are indicated on top of the figure, separated by dashed lines.
 }
\label{Fig:numStars}
\end{center}
\end{figure}

The fraction of main-sequence stars in NGC~3766 that are detected to be periodically variable is shown in Fig.~\ref{Fig:numStars}.
It is about 50\% at magnitudes between 11 and 11.5~mag, i.e. at the lower luminosity end of SPB stars (our data unfortunately cannot tell the number of SPB stars at higher luminosities, due to photometric saturation of the bright stars in the survey).
In the region between SPB and $\delta$~Sct stars, where the new variables are found, this fraction decreases continuously from $\sim$40\% at 11.5~mag to $\sim$10\% at 13.5~mag.
This is of interest in the exploration of the possible origin of the variability.

\section{Possible origins of the periodic variability}
\label{Sect:origin}

Three scenarios can be thought of to explain periodic photometric variability of B- and A-type main-sequence stars: pulsation, stellar rotation (spotted stars), and non-sphericity due to binarity (ellipsoidals).
They are reviewed in \cite[Mowlavi et al.~(2014)]{Mowlavi_etal14} in relation to the new periodic variables found in NGC~3766.
None of those scenarios can currently provide a satisfactory explanation for the observations of those stars, though:
pulsation is not predicted in this region of the main sequence by \textit{standard} pulsation models;
spots at the surface of rotating stars are not predicted in late B- and early A-type stars;
and binarity, at the origin of ellipsoidal systems, cannot explain multiperiodicity.

Of relevance to this conference is the question whether a pulsation origin is a viable scenario for those periodic stars, despite the fact that they lie in the pulsation `gap' on the main sequence.
It is interesting to note in this respect that four of those periodic variables for which spectra are available in the literature are rotating faster than half their critical equatorial velocity (see MBSE13).
While small number statistics, it suggests that fast rotation may be a key issue in understanding their origin.

No pulsation model prediction exists yet for very rapidly rotating stars.
However, models with rotational velocities approaching half the critical velocity do predict pulsation frequencies significantly different than the ones predicted for non-rotating stars (\cite[Townsend 2005]{Townsend05}, \cite[Ushomirsky \& Bildsten 1998]{UshomirskyBildsten98}).
They also show that rapid rotation can excite modes that are damped in the absence of rotation, opening the possibility to find pulsation outside the classical instability strips.
Interestingly, the fraction of stars that are periodic variables reaches a maximum close to the region of SPB stars, and decreases for fainter stars as shown in Sect.~\ref{Sect:properties}.
This may be a further clue supporting pulsation in rapidly rotating stars.

\section{Potential contribution of Gaia}
\label{Sect:gaia}

The understanding of the new periodic B- and A-type variables would benefit from an all-sky census.
The space-based Gaia mission\footnote{See http://www.rssd.esa.int/index.php?project=GAIA}, to be launched towards the end of 2013, is potentially a very good candidate to achieve this, for several reasons:
\begin{itemize}
\item[--]It is a multi-epoch mission, with a mean of about 75 measurements per star over the 5-year mission duration.
\item[--]The photometric uncertainty equals about 1~mmag up to 14~mag in $G$, the main Gaia photometric band (see Fig.~19 of \cite[Jordi et al.~2010]{Jordi_etal10}).
\item[--]The scanning law leads to a time sampling well suited to detect periodic variables with periods below several days.
The spectral window presents only few peaks, that are unlikely to affect the frequency range of interest (below 10~d$^{-1}$) when convoluted with a periodic signal (see Fig.~1 of \cite[Eyer et al.~2009]{Eyer_etal09}).
As a result, the expected period recovery rate is very good for the periods typical of the new periodic variables, even for low signal-to-noise ratios.
Assuming an uncertainty (Gaussian noise) of 1~mmag in the measurement, \cite[Eyer \& Mignard (2005)]{EyerMignard05} predict a period recovery rate above 85\% for a sinusoidal signal of amplitude 1.1~mmag and a frequency of 3~d$^{-1}$, and above 95\% for an amplitude of 1.3~mmag.
\item[--]Gaia will provide parallaxes with unprecedented accuracy, allowing the positioning of the stars in the H-R diagram with high accuracy.
\item[--]Gaia's photometric instrument includes spectro-photometric measurements in two color bands.
In addition, Gaia also includes a spectroscopic instrument.
From those data, astrophysical information such as atmospheric parameters and rotational velocities will be derived for stars brighter than about 12~mag\footnote{From http://www.rssd.esa.int/index.php?project=GAIA\&page=Science\_Performance (estimate as of 09/2013).} in $G_\mathrm{RVS}$, the band of Gaia's spectrometer.
This possibility of Gaia to characterize the nature of bright stars is a great advantage to explore the nature of periodic stars.
\item[--]Finally, Gaia is an all-sky survey.
\end{itemize}

\section{Conclusions}
\label{Sect:conclusions}

The discovery in NGC~3766 of late B- and early A-type stars that show periodic variability in their light curve provides a challenge for stellar physics.
Neither pulsation is expected in those stars, because they lie in the pulsation `gap' on the main sequence, nor spots on their surface, because they have too large effective temperatures.

Not all thirty-six stars in this new class of periodic variables need necessarily to be of the same type of variability, but pulsation in a rapidly rotating star may, we believe, be a good scenario for a fraction of them.
Which fraction is a question that must be addressed by further observations, that we have already planned to undertake.


%
%

\end{document}